\documentclass[12pt]{iopart}
\usepackage{graphicx}
\usepackage{amssymb,mathptmx}
\usepackage[colorlinks=true,citecolor=blue,linkcolor=blue,urlcolor=blue,bookmarks=false]{hyperref}
\usepackage{gensymb}

\usepackage{tikz}
\usetikzlibrary{arrows,snakes,positioning,shapes,patterns}

\def\unit#1{\mathord{\thinspace\rm #1}}
\def\func#1{\mathop{\rm #1}\nolimits}

\begin{document}

\title{Quantum capacitive coupling between large-angle twisted graphene layers}

\author{Alina Mre\'nca-Kolasi\'nska$^{1,2}$, Peter Rickhaus$^3$, Giulia Zheng$^3$, Klaus Richter$^4$, Thomas Ihn$^3$, Klaus Ensslin$^3$ and Ming-Hao Liu$^1$}

\address{$^1$ Department of Physics, National Cheng Kung University, Tainan 70101, Taiwan}
\address{$^2$ AGH University of Science and Technology, Faculty of Physics and Applied Computer Science, al. Mickiewicza 30, 30-059 Krak\'ow, Poland}
\address{$^3$ Solid State Physics Laboratory, ETH Z\"urich, CH-8093 Z\"urich, Switzerland}
\address{$^4$ Institut f\"ur Theoretische Physik, Universit\"at Regensburg, D-93040 Regensburg, Germany}
\ead{ \mailto{minghao.liu@phys.ncku.edu.tw}, \mailto{alina.mrenca@fis.agh.edu.pl}}

\date{\today}

\begin{abstract}

Large-angle twisted bilayer graphene (tBLG) is known to be electronically decoupled due to the spatial separation of the Dirac cones corresponding to individual graphene layers in the reciprocal space. 
The close spacing between the layers causes strong capacitive coupling, opening possibilities for applications in atomically thin devices. 
Here, we present a self-consistent quantum capacitance model for the electrostatics of decoupled graphene layers, and further generalize it to deal with decoupled tBLG at finite magnetic field and large-angle twisted double bilayer graphene at zero magnetic field. We probe the capacitive coupling through the conductance, showing good agreement between simulations and experiments for all the systems considered. We also propose a new experiment utilizing the decoupling effect to induce a huge and tunable bandgap in bilayer graphene by applying a moderately
low bias. Our model can be extended to systems composed of decoupled graphene multilayers as well as non-graphene systems, opening a new realm of quantum-capacitively coupled materials.
\end{abstract}
\noindent{\it Keywords\/}: twisted bilayer graphene, double twisted bilayer graphene, quantum capacitance

\maketitle

\begin{figure}[b]
\includegraphics[width=\columnwidth]{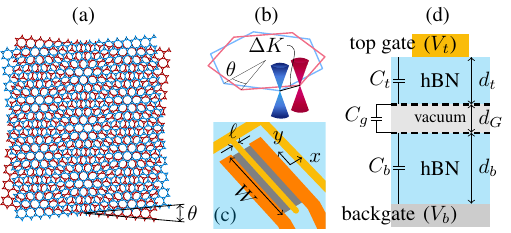} 
\caption{\textbf{The tBLG device design.} Schematics of (a) a tBLG lattice composed of two graphene layers twisted by an angle $\theta$ and (b) their corresponding Dirac cones in reciprocal space. The dual-gated two-terminal decoupled tBLG device considered in the transport experiment and simulations is sketched in (c) for a perspective top view and (d) for its side view.} 
\label{fig:TBG_B0}
\end{figure}

Recently, there has been an increasing interest in thin van der Waals heterostructures \cite{Geim2013, Kim2016tearandstack}, including twisted bilayer graphene (tBLG). 
In tBLG [figure~\ref{fig:TBG_B0}(a)], the Brillouin zones of the two layers are rotated against each other [figure~\ref{fig:TBG_B0}(b)], and a large twist angle leads to the separation of the Dirac cones of both layers 
\cite{Fallahazad2008, Luican2011, Sanchez2012qhe, Sanchez-Yamagishi2017, Rickhaus2020electronicthickness}. This suppresses interlayer scattering due to the  large momentum difference, making the two layers essentially electronically decoupled \cite{Schmidt2008decoupled, Schmidt2010decoupled, Kim2013tbg_coherence, Deng2020decoupling, piccinini2021parallel}. However, their atomically thin layer spacing allows them to couple electrostatically because the electric charge on one layer causes an effective gating of the other layer. 
This mechanism enables realization of atomically thin devices composed of decoupled layers, with the large twist being an alternative to isolating the layers with dielectrics \cite{Kim2012, Gorbachev2012CoulombDrag, Li2017, Lee2017, Liu2019FQHE}. 
However, the strong quantum capacitive coupling makes precise electrostatic modeling indispensable for simulation of these devices \cite{Fang2007, Xia2009}.

In this work, we present the self-consistent quantum capacitance model used in reference\ \cite{Rickhaus2020electronicthickness} for decoupled tBLG at zero magnetic field and generalize it considerably to deal with decoupled tBLG in the presence of magnetic field, decoupled twisted double Bernal-stacked bilayer graphene (tdBLG), and decoupled multilayer graphene systems. The quantum conductance of such layered structures depends strongly on the capacitive coupling and can be used as a sensitive probe of the latter. We show quantitatively good agreement with our own experimental results for a dual-gated two-terminal tBLG device sketched in Figs.\ \ref{fig:TBG_B0}(c) and (d), showing strong reliability of our model. For tdBLG, our transport simulations agree well with the experimental findings \cite{Rickhaus2019tdblg}, despite the strong complication due to the gate-tunable band gap \cite{McCann2006, Castro2007gap, Oostinga2008gap, Zhang2009gap, Mak2009gap, Taychatanapat2010gap}. We also propose employing the decoupling mechanism to build a thin capacitor composed of three Bernal-stacked bilayers (BLGs) brought close to each other at a large rotation angle, where the outer ones play the role of the electrostatic gates. Thanks to the large capacitance between the BLGs, and a resulting large bias, it is possible to induce a large bandgap in the middle BLG and investigate the transport features occurring close to the band edge.
Our models can be in general applied to electronically decoupled materials that are quantum-capacitively coupled to each other, including topological insulator surface states \cite{Ziegler2020hgte}, but is not limited to alike layers, being adaptable to hybrid systems consisting of different materials hosting two-dimensional electron gas \cite{Simonet2015}.

\section{Methods}
\paragraph*{Self-consistent quantum capacitance model for tBLG.}

To model the decoupled tBLG device, we assume two layers of graphene described by the linear Dirac dispersion relation $E=\pm \hbar v_{\mathrm{F}} k$, where $\hbar$ is the reduced Planck constant and $v_{\mathrm{F}}\approx 10^6\unit{m\cdot s^{-1}}$ is the Fermi velocity of graphene. We adopt $\hbar v_{\mathrm{F}} \approx 3\sqrt{3}/8\unit{eV\cdot nm}$. The two electronically decoupled single-layer graphene (SLG) flakes are tightly spaced (assuming the spacing to be $d_{\mathrm{G}}=0.12\unit{nm}$ found in reference\ \cite{Rickhaus2020electronicthickness}) such that a tiny shift of the Fermi energy of the first layer causes an appreciable gating effect on the second layer, which in turn acts as a back gate of the first one. The whole process is iterated using the formulas derived in \cite{Liu2013multigated}. The carrier density
\begin{equation}
n = n_{\mathrm{C}}+\Delta n\ 
\label{eq:n}
\end{equation}
of a SLG free of intrinsic doping and subject to two gates at voltages $V_1$ and $V_2$ is composed of the classical carrier density
\begin{equation}
n_{\mathrm{C}} = \sum_{i=1,2}\frac{C_{i\mathrm{G}}}{e}V_i\ ,
\label{eq:nC}
\end{equation}
where $e>0$ is the elementary charge and $C_{i\mathrm{G}}$ is the capacitance (per unit area) between gate $i$ and graphene, and the correction
\begin{equation}
\Delta n = \func{sgn}(n_{\mathrm{C}}) n_{\mathrm{Q}} \left(1- \sqrt{1+2\frac{|n_{\mathrm{C}}|}{n_{\mathrm{Q}} }}\right)
\label{eq:Delta n}
\end{equation}
accounting for the quantum nature of the finite density of states of the conducting plate, where
\begin{equation}
n_{\mathrm{Q}} = \frac{\pi}{2} \left(\frac{\hbar v_{\mathrm{F}}}{e} \frac{C_\mathrm{1G}+C_\mathrm{2G}}{e}\right)^2
\label{eq:nQ}
\end{equation}
arizes solely from the quantum capacitance \cite{Luryi1988,Fang2007}. The corresponding electric potential of the graphene sheet is given by
\begin{equation}
V_\mathrm{G} = -\frac{e\Delta n}{C_\mathrm{1G}+C_\mathrm{2G}}\ .
\label{eq:VG}
\end{equation}
To apply Eqs.\ \ref{eq:n}--\ref{eq:VG} to the dual-gated decoupled tBLG device sketched in figure~\ref{fig:TBG_B0}(d), we consider the top graphene layer (upper dashed line) to be dual-gated by the top gate at voltage $V_\mathrm{t}$ and bottom graphene layer (lower dashed line) at electric potential $V_\mathrm{Gb}$. Substituting $V_1=V_\mathrm{t}$, $V_2=V_\mathrm{Gb}$, $C_\mathrm{1G}=C_\mathrm{t}$, and $C_\mathrm{2G}=C_\mathrm{g}$ into Eqs.\ \ref{eq:nC}--\ref{eq:VG}, we obtain the electric potential $V_\mathrm{Gt}$ of the top graphene layer, given $V_\mathrm{t}$ and $V_\mathrm{Gb}$, the former being a fixed input while the latter to be self-consistently iterated. Similarly, the bottom graphene layer is dual-gated by the top graphene layer at potential $V_\mathrm{Gt}$ and back gate at voltage $V_\mathrm{b}$. Substituting $V_1=V_\mathrm{Gt}$, $V_2=V_\mathrm{b}$, $C_{1G}=C_\mathrm{g}$, and $C_\mathrm{2G}=C_\mathrm{b}$ into Eqs.\ \ref{eq:nC}--\ref{eq:VG}, we obtain the electric potential $V_\mathrm{Gb}$ of the bottom graphene layer, given $V_\mathrm{Gt}$ and $V_\mathrm{b}$, the former being just computed and the latter being a fixed input. The newly obtained $V_\mathrm{Gb}$ is used to compute $V_\mathrm{Gt}$ and vice versa iteratively, until $V_\mathrm{Gt}$ and $V_\mathrm{Gb}$ both converge to a satisfactory precision. From the converged $V_\mathrm{Gt}$ and $V_\mathrm{Gb}$ one obtains the individual carrier density $n_\mathrm{t}$ ($n_\mathrm{b}$) for the top (bottom) layer, using Eqs.\ \ref{eq:n}--\ref{eq:nQ}. Then, $-eV_\mathrm{Gt}$ and $-eV_\mathrm{Gb}$ are the onsite energies entering the tight-binding Hamiltonian for transport calculations to be explained later. 
The capacitance and intrinsic doping can in general be position-depent, $C_\mathrm{t}=C_\mathrm{t}(x)$, and the convergence is required for all $x$, leading all $n_\mathrm{t},n_\mathrm{b},V_\mathrm{Gt},V_\mathrm{Gb}$ to depend on $x$. For capacitance $C=C(x,y)$ the same requirement applies for all relevant $x$ and $y$, but in each case the iteration converges rapidly. 

For strong magnetic field we account for the Landau quantization of the density of states, 
$
D(E,B_z) = \frac{4e B_z}{h} \sum\limits_{n_\mathrm{L}} \delta(E - E_{n_\mathrm{L}}),
$
where $E_{n_\mathrm{L}}=\mathrm{sgn}(n_\mathrm{L}) \sqrt{2e B_z \hbar v_{\mathrm{F}}^2|n_\mathrm{L}|}$ and $n_\mathrm{L}=0,\pm 1, \pm 2, \dots$. 
 The carrier density is given by
\begin{equation} 
n(E,B_z) = \int_0^E D(E', B_z) \rmd E'.
\label{eq:dens_int}
\end{equation} 
To account for the Landau level (LL) broadening, we approximate the Dirac delta by a Lorentzian function, and the integration (\ref{eq:dens_int}) can be done analytically. The resulting carrier density is quantized in energy and magnetic field. 
The carrier density given by equation~(\ref{eq:dens_int}) is equal to the sum of the gate-induced doping 
\begin{equation}
\label{eq:LL_eq} 
n(eV_\mathrm{G}, B_z) = \frac{C_\mathrm{1G}}{e} (V_1 - V_\mathrm{G})+ \frac{C_\mathrm{2G}}{e} (V_2 - V_\mathrm{G}).
\end{equation} 
We solve equation\ (\ref{eq:LL_eq}) for $V_\mathrm{G}$ numerically. 
Then, for two decoupled graphene layers in strong external magnetic field, the calculation of the electric potential $V_\mathrm{Gt}$ and $V_\mathrm{Gb}$ is done in a similar self-consistent way as for the linear dispersion relation 
but substituting eqations~(\ref{eq:n})--(\ref{eq:VG}) with the numerical solution of equation~(\ref{eq:LL_eq}) (for details see  the Supplemental Materials \cite{supplement}).

\paragraph*{Quantum capacitance model for tdBLG.}
We first develop a quantum capacitance model for an individual dual-gated BLG based on Refs.~\cite{McCann2013, tomadin2021}. We consider a general case with the onsite energy given by $U_0 \pm U / 2 $ for the top (bottom) layer, where $U$ is the asymmetry parameter and $U_0=-eV_\mathrm{G}$ is the band offset.

The relation between the carrier density, the asymmetry parameter, and the band offset 
is given by \cite{Varlet2014blg}
\begin{equation}
-eV_\mathrm{G} = -\mathrm{sgn}(n) \sqrt{ \frac{\gamma_1^2}{2} + \frac{U^2}{4} + \hbar^2 v_{\mathrm{F}}^2 \pi |n| - \frac{\gamma_1}{2} \sqrt{
\gamma_1^2 + (2\hbar v_{\mathrm{F}})^2\pi |n| \left(1+ \frac{U^2}{\gamma_1^2} \right)} }.
\label{eq:bandstruct}
\end{equation} 
For a dual gated sample, and the densities of the top and bottom layer given by $n_\mathrm{t}$ and $n_\mathrm{b}$, respectively, we obtain 
\begin{eqnarray}
n_\mathrm{b} - n_\mathrm{t} =& -\frac{n_{\perp} U}{2\gamma_1} 
\ln\left(
\frac{|n|}{2 n_\perp} + \frac{1}{2} \sqrt{ \left(\frac{n}{n_\perp} \right)^2 + 
\left(\frac{U}{2\gamma_1} \right)^2} 
\right), \\
n_\mathrm{b} - n_\mathrm{t} =& \frac{C_\mathrm{b}}{e} \left[V_\mathrm{b} - \left(V_\mathrm{G} + \frac{U}{2e} \right)\right]
- \frac{C_\mathrm{t}}{e} \left[V_\mathrm{t} - \left(V_\mathrm{G} - \frac{U}{2e} \right)\right] - \frac{2C_\mathrm{g}}{e} \frac{U}{e},\\
\label{eq:nsum}
n = n_\mathrm{b} + n_\mathrm{t} =& \frac{C_\mathrm{b}}{e} V_\mathrm{b} + \frac{C_\mathrm{t}}{e} V_\mathrm{t}- \frac{C_\mathrm{b} + C_\mathrm{t}}{e}V_\mathrm{G} +\frac{C_\mathrm{t} - C_\mathrm{b}}{e}\frac{U}{2e},
\end{eqnarray}
where $\gamma_1=0.39$ eV is the nearest-neighbor hopping for the interlayer coupling, and $n_{\perp} = \gamma_1^2/\pi\hbar^2 v_{\mathrm{F}}^2$ is the characteristic carrier density. For the details of the derivation see reference~\cite{supplement}. The system of four nonlinear equations (\ref{eq:bandstruct} -- \ref{eq:nsum}) is solved numerically to obtain $\Delta n = n_\mathrm{t}-n_\mathrm{b}$, $n=n_\mathrm{t}+n_\mathrm{b}$, $U$, and $V_\mathrm{G}$. 

With the quantum capacitance model for BLG at hand, we can further extend it to tdBLG. To this end, we consider two stacked BLGs, coupled capacitively to external top and bottom gates and to each other [see figure~\ref{fig:tdBLG}(c)]. 
The problem can be then solved self-consistently as for tBLG; however, we found treating the problem as a set of eight coupled nonlinear equations more efficient (see \cite{supplement}).

\paragraph*{Quantum transport calculation.}
To simulate real devices and speed up calculations, the hopping parameter $t_0$ and lattice spacing $a_0$ approximated by $3\unit{eV}$ and $1/4\sqrt{3}\unit{nm}$, respectively, are scaled to $t_0/s_{\mathrm{f}}$ and $s_{\mathrm{f}} a_0$ \cite{2015scaling} using $s_{\mathrm{f}}=4$ (only for BLG and tdBLG we use $s_{\mathrm{f}}=2$).
We express the two-terminal conductance as $G=(R_\mathrm{c}+G_0^{-1})^{-1}$, where $R_\mathrm{c}$ is the contact resistance and $G_0$ is the ballistic conductance calculated using the real-space Green's function approach \cite{Datta1995}. 
For tBLG devices at zero or low magnetic field, we use $R_\mathrm{c}=0.005\unit{h/e^2}$ as a reasonable parameter for the contact resistance.
Here, as well as for tdBLG we assume the system is translationally invariant along the lateral direction, and $G_0$ is computed by using the method of periodic boundary hopping \cite{Wimmer2008,2012periodicModel, Liu2012periodic}: $G_0 = (W/3\pi s_{\mathrm{f}} a_0)(g_\mathrm{b}+g_\mathrm{t})$, where $g_j=(e^2/h)\int_{-k_{\mathrm{F}}}^{k_{\mathrm{F}}}T(k_y)\rmd k_y$ ($k_{\mathrm{F}}$ being the Fermi momentum) \cite{Kang2020} is the normalized conductance of the bottom (top) graphene layer for $j=b$ ($j=t$). The system is described by the tight-binding Hamiltonian
\begin{equation}
H_j = H_0-e\sum_n V_{\mathrm{G}j}(x_n)c_n^\dag c_n\ ,
\label{eq:H}
\end{equation}
where $j=t,b$ is the layer index, $H_0$ is the clean part of the minimal tight-binding model for bulk graphene \cite{2012periodicModel, Liu2012periodic}, and the operator $c_n$ ($c_n^\dag$) annihilates (creates) an electron on site $n$ located at $(x_n,y_n)$. It is the second term in equation\ (\ref{eq:H}) for the onsite energy where the electric potential $V_\mathrm{Gb}$ and $V_\mathrm{Gt}$, found from the self-consistent electrostatic model, enter the transport calculations. 

For the transport modeling of tBLG at large magnetic field we use the wave-function matching method \cite{Kol2016transport}
for graphene, considering a zigzag ribbon of width 400 nm. 
At zero-temperature the conductance is calculated using the Landauer formula $G(B)=2e^2 T(B)/h $, with $T(B)$ being the transmission summed over all modes.

\section{Results}
\paragraph*{Self-consistent quantum capacitance model for decoupled tBLG.}

Reference \cite{Rickhaus2020electronicthickness} investigated dual-gated two-terminal devices consisting of decoupled large-angle tBLG samples, schematically shown in figure~\ref{fig:TBG_B0}(c) for a perspective top view and in figure~\ref{fig:TBG_B0}(d) for its side view. In the following discussion we focus on a device fabricated with a top gate of length $\ell=320\unit{nm}$ and sample width $W\approx 2.9\unit{\mu m}$. Details of the device fabrications are given in reference\ \cite{Rickhaus2020electronicthickness}.

\paragraph*{Decoupled tBLG without magnetic field.}\label{sec:TBG0}

\begin{figure}[t]
\includegraphics[width=\columnwidth]{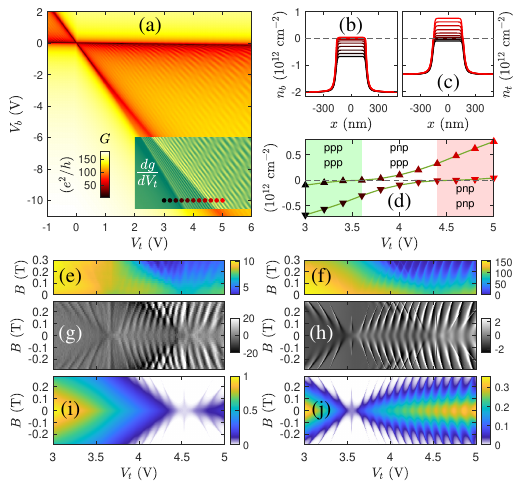}
\caption{\textbf{$B=0$ to low $B$.} (a) Calculated two-terminal conductance $G$ as a function of top gate voltage $V_\mathrm{t}$ and back gate voltage $V_\mathrm{b}$ at zero magnetic field $B=0$. The overlay at the bottom right corner shows the numerical derivative of the normalized conductance $g$ with respect to $V_\mathrm{t}$. Carrier density profiles $n_\mathrm{b}(x)$ and $n_\mathrm{t}(x)$ of the bottom and top graphene layer are shown in (b) and (c), respectively, subject to gate voltage configurations marked in (a) with the line and dot colors corresponding to each other. (d) $n_\mathrm{b}(x=0)$ and $n_\mathrm{t}(x=0)$ marked by $\bigtriangledown$ and $\bigtriangleup$ as a function of $V_\mathrm{t}$ at fixed $V_\mathrm{b}=-10\unit{V}$ corresponding to panel (b) and (c), respectively. Shaded areas distinguish three regions: both layers in ppp (light green), top layer in pnp but bottom layer in ppp (white), and both layers in pnp (pink). (e) Measured and (f) simulated two-terminal conductance $G$ as functions of $V_\mathrm{t}$ and $B$ up to $0.3\unit{T}$. Fabry-P\'erot interference fringes of $\rmd G/\rmd V_\mathrm{t}$ from the experiment and $\rmd g/\rmd V_\mathrm{t}$ from the simulations are shown in (g) and (h), respectively. Calculated normalized conductance $g$ for the (i) bottom and (j) top graphene layer. Color bars are in units of $e^2/h$ for (e), (f), (i), and (j), and $e^2/h \unit{V^{-1}}$ for (g) and (h).}
\label{fig:tBLG zero and weak B}
\end{figure}

We assume the two layers of graphene without magnetic field are described by the linear Dirac dispersion relation.
\autoref{fig:tBLG zero and weak B}(a) shows the computed two-terminal conductance simulated for the considered decoupled tBLG device sketched in Figs.\ \ref{fig:TBG_B0}(c)--(d) as a function of $V_\mathrm{t}$ and $V_\mathrm{b}$. The diagonal charge neutrality line splits into two which is a signature of the decoupling of the two graphene layers. The splitting as well as the superimposed Fabry-P\'erot (FP) interference fringes \cite{Young2009, Rickhaus2013ballisticInterferences} are better seen by mean of numerical differentiation. We show $\rmd g/\rmd V_\mathrm{t}$ as an overlaid inset on figure~\ref{fig:tBLG zero and weak B}(a), where the horizontally aligned dots mark the scan with $3\unit{V}\leq V_\mathrm{t}\leq 5\unit{V}$ at $V_\mathrm{b}=-10\unit{V}$ that we are going to focus on for the rest of the discussions of the decoupled tBLG device. Along this $V_\mathrm{t}$ scan, the carrier density profiles $n_\mathrm{b}(x)$ and $n_\mathrm{t}(x)$ are shown in figure\ \ref{fig:tBLG zero and weak B}(b) and (c), respectively. The $V_\mathrm{t}$ dependence of $n_\mathrm{b}(0)$ and $n_\mathrm{t}(0)$ is shown in figure\ \ref{fig:tBLG zero and weak B}(d), where three regions can be clearly seen: Both graphene layers in unipolar ppp for $V_\mathrm{t}\lesssim 3.6\unit{V}$, top layer in pnp but bottom layer remaining in ppp for $3.6\unit{V}\lesssim V_\mathrm{t}\lesssim 4.5\unit{V}$, and both layers in pnp for $V_\mathrm{t}\gtrsim 4.5\unit{V}$. These regions are characterized by no FP fringes, one set of FP fringes and two sets of FP fringes, respectively.

\paragraph*{Decoupled tBLG with magnetic field.} We next go beyond reference\ \cite{Rickhaus2020electronicthickness} to account for magnetotransport in the same decoupled tBLG device, where the uniform magnetic field $B$ is applied along $z$ perpendicular to the graphene layers. When $B$ is weak, the Dirac linear dispersion remains valid, and the above introduced self-consistent model can be directly applied. \autoref{fig:tBLG zero and weak B}(e) shows the experimentally measured two-terminal conductance $G$ as a function of $V_\mathrm{t}$ restricted to the range marked in figure\ \ref{fig:tBLG zero and weak B}(a) and $B$ up to $0.3\unit{T}$. Our simulated $G$ shown in figure\ \ref{fig:tBLG zero and weak B}(f) exhibits a similar profile, despite the different magnitude of $G$. To better compare the details with our simulation, we mirror the experimental data about the $V_\mathrm{t}$ axis and show $\rmd G/\rmd V_\mathrm{t}$ in figure\ \ref{fig:tBLG zero and weak B}(g). It exhibits complex FP fringes that are satisfactorily consistent with our computed differentiated normalized conductance $\rmd g/\rmd V_\mathrm{t}$ shown in figure\ \ref{fig:tBLG zero and weak B}(h). Closer inspection of the region with $V_\mathrm{t}\gtrsim 4.5\unit{V}$ shows that there are two sets of FP fringes superimposed, one dispersing with $B$ slower and the other faster. The slower (faster) set is expected to come from the top (bottom) graphene layer because of the higher (lower) gating efficiency; the layer with lower gating efficiency needs a larger gate voltage to compete with the $B$-dependent Aharanov-Bohm phase picked up by the interfering electron within the FP cavity \cite{Rickhaus2015}. This is confirmed by showing the individual contribution $g_\mathrm{b}$ and $g_\mathrm{t}$ in figure\ \ref{fig:tBLG zero and weak B}(i) and (j), respectively, which sum up to $g=g_\mathrm{b}+g_\mathrm{t}$.

\begin{figure}[b]
\includegraphics[width=\columnwidth]{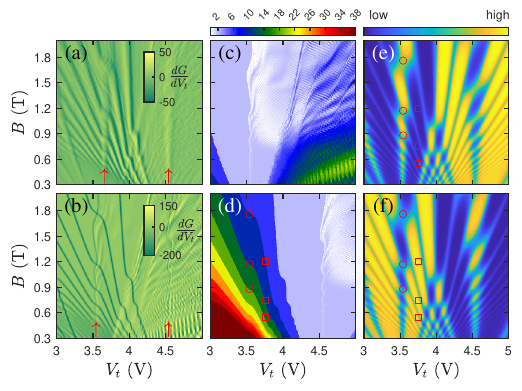}
\caption{\textbf{Strong magnetic field transport in tBLG.} Numerical derivative of the (a) measured and (b) calculated two-terminal conductance as a function of
top gate voltage $V_\mathrm{t}$ and magnetic field at $V_\mathrm{b}=-10$ V. (c), (d) Individual conductances calculated for the top and bottom layer, respectively. (e) $\rmd n_\mathrm{t}/\rmd V_\mathrm{t}$ and (f) $\rmd n_\mathrm{b}/\rmd V_\mathrm{t}$ values show a clear correspondence to the conductance plateaus kinks.} \label{fig:grad_LL}
\end{figure}

Next we turn to strong external magnetic field.
Figure \ref{fig:grad_LL}(a) shows the 
experimentally measured and figure~\ref{fig:grad_LL}(b) the calculated transconductance obtained at
$V_\mathrm{b} =-10$ V as a function of the top-gate voltage and magnetic field. In the transconductance map, two sets of Landau levels  are visible, emerging from the two split charge neutrality points, marked by red arrows in figure~\ref{fig:grad_LL}(a) and (b). The individual top and bottom layer conductances are shown in figure~\ref{fig:grad_LL}(c) and (d), respectively, confirming that the entire LL spectrum consists of two layers' superimposed Landau fans, and that the layers remain electrically decoupled at strong magnetic field.
The Landau fans are dramatically different from the commonly observed ones in graphene,
and exhibit ''kinks'' at the crossing between the Landau levels of the two layers. 
Their origin can be understood by comparing the conductance map to the top and bottom layer density gradient with respect to $V_\mathrm{t}$ in figure~\ref{fig:grad_LL}(e), and (f), respectively.

We first focus on the 2nd, 3rd, and 4th LL of the bottom layer marked by squares in figure~\ref{fig:grad_LL}(d).
The density of states (DoS) is high at the LL, as is the density per gate voltage. Thus, the $\rmd n_\mathrm{b}/\rmd V_\mathrm{t}$ value is high, [figure~\ref{fig:grad_LL}(f)] and the $\rmd n_\mathrm{t}/\rmd V_\mathrm{t}$ value [figure~\ref{fig:grad_LL}(e)] is low as the total carrier density induced by the top gate is conserved.
On the other hand, the points marked with circles in figure~\ref{fig:grad_LL}(d), (e) and (f) are along the top layer 0th LL, and the top layer DoS is high. Based on the argument above, here the $\rmd n_\mathrm{t}/\rmd V_\mathrm{t}$ ($\rmd n_\mathrm{b}/\rmd V_\mathrm{t}$) value is high (low). Therefore, we expect the slope of the LLs to change, and in particular at the points marked with circles, the bottom layer LLs slope becomes smaller. 
Recent experiment \cite{piccinini2021parallel} reported similar effects.
Note that this feature is only recovered in the LL-quantized-density model. For the result obtained with the linear dispersion relation see \cite{supplement}. 
The good qualitative agreement between the experimental and theoretical results shows that the self-consistent model is accurate for other than linear dispersion relations.

\begin{figure}[t]
\includegraphics[width=\columnwidth]{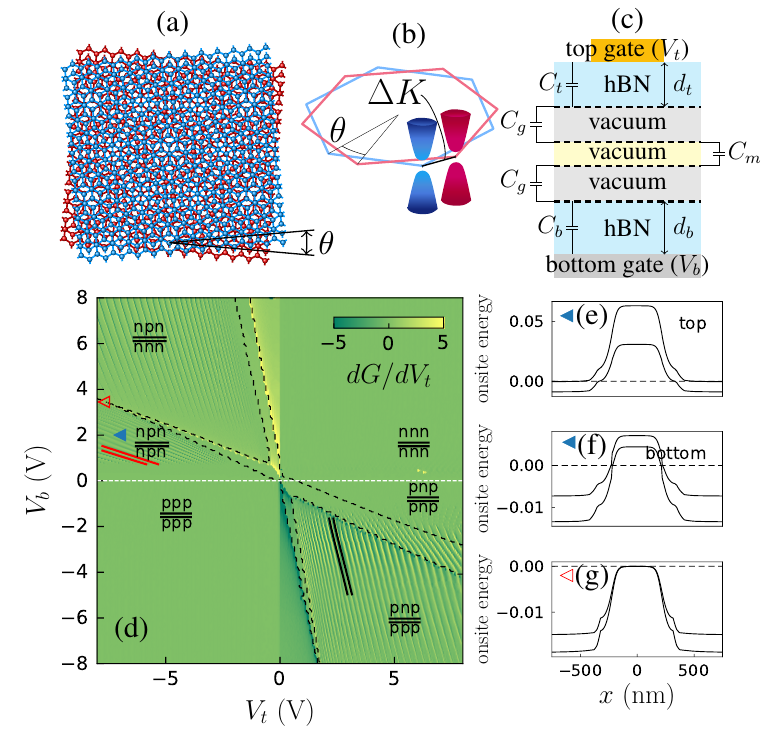}
  \caption{\textbf{Fabry-P\'erot oscillation in tdBLG.} (a) Sketch of the Bernal stacked large-angle tdBLG and (b) its low-energy bandstructure. 
  (c) Schematic of the dual-gated tdBLG system. (d) Numerical derivative of two-terminal conductance as a function of top gate and back gate voltage in tdBLG. The onsite energy profiles at the voltage configuration marked by $\triangleleft$ are shown in (e) for the top BLG and (f) and (g) for the bottom BLG.
  } \label{fig:tdBLG}
\end{figure}

\paragraph*{Decoupled tdBLG.}
We next consider large-angle tdBLG [figure~\ref{fig:tdBLG}(a)], where,
similar to the tBLG case, the two BLGs are decoupled electronically by the large momentum separation of the Dirac cones of the two BLGs [figure~\ref{fig:tdBLG}(b)]. 

The gate capacitances $C_\mathrm{t}$, $C_\mathrm{b}$ are obtained from a finite element electrostatic simulation for a sample geometry adopted 
from reference~\cite{Rickhaus2019tdblg}, with the hBN thicknesses $d_\mathrm{b}=90$ nm, $d_\mathrm{t}=60$ nm, placed between a global back gate and narrow top gate width 400 nm. 
The interlayer capacitance within an individual BLG is assumed to be
$C_\mathrm{g} = 7.4\ \mu \mathrm{Fcm^{-2}}$ 
\cite{Rickhaus2020electronicthickness}, whereas the value of the capacitance between the BLG layers is $C_\mathrm{m}=3.5\ \mu \mathrm{Fcm^{-2}}$ \cite{Rickhaus2019tdblg}.

In the self-consistent model for tdBLG, we include the effect of the crystal field \cite{Gruneis2008, Haddadi2020, Tepliakov2021crystalfield, Rickhaus2021tdblg} which was shown in reference~\cite{Rickhaus2019tdblg} to open a bandgap even without gate voltage. In the tdBLG sample the inner and outer graphene layers see a different environment, and thus feel an unequal electrostatic potential, which effectively creates an intrinsic bias. This induces a small negative charge in the inner layers. We can include this effect in our model by assuming a constant density difference $\Delta n_{0}$ between the bottom and top layer of a BLG. From the measured values of the displacement field needed to close the bandgaps \cite{Rickhaus2019tdblg} we estimate $\Delta n_{0,1}=13\times 10^{11} \mathrm{cm}^{-2}$ for the upper BLG and  $\Delta n_{0,2}=-14\times 10^{11} \mathrm{cm}^{-2}$ for the lower BLG. 

Figure \ref{fig:tdBLG}(d) shows the transconductance as a function of $V_\mathrm{t}$ and $V_\mathrm{b}$, which recovers the key features observed in reference~\cite{Rickhaus2019tdblg}. The FP oscillations occur when a bipolar junction is formed in the top or bottom BLG. Interestingly, for negative $V_\mathrm{b}$ the oscillations occur for the top BLG only (the slope highlighted by black solid lines). Conversely, for $0<V_\mathrm{b}<4$ V the FP oscillation is present only for the bottom BLG (highlighted by red lines); at higher $V_\mathrm{b}$ only a faint oscillation for the top BLG can be spotted, when the n-p-n junction is formed in the upper BLG [see labels in figure~\ref{fig:tdBLG}(d)]. 
This difference in the visibility of the oscillations can be explained as due to a large bandgap across the device, which strongly reduces the transmission through the cavity when the p-n interface is smooth. For example, the oscillation in the top BLG is hardly visible at $V_\mathrm{b}>0$ where the bandgap at the p-n interface happens to be large [see the onsite energy profile in figure~\ref{fig:tdBLG} (e)], but in the bottom BLG the bandgap is reduced by the applied displacement field [figure~\ref{fig:tdBLG} (f)]. 

Another feature which our model captures in good qualitative agreement with experiment is the bandgaps in the top-gated region that are opened even at low applied gate voltages [shown in figure~\ref{fig:tdBLG}(d) by black dashed lines] and closed at  $(2.2, -10.6)$V for the upper BLG and  $(V_\mathrm{t}, V_\mathrm{b})\approx(-7.7,3.4)$V for the lower BLG [see figure~\ref{fig:tdBLG} (g)].  
A feature not taken into account by the model is the difference in the electron and hole effective mass leading to a kink of the charge neutrality line in the experiment \cite{Rickhaus2019tdblg}.

\begin{figure}[t]
\includegraphics[width=\columnwidth]{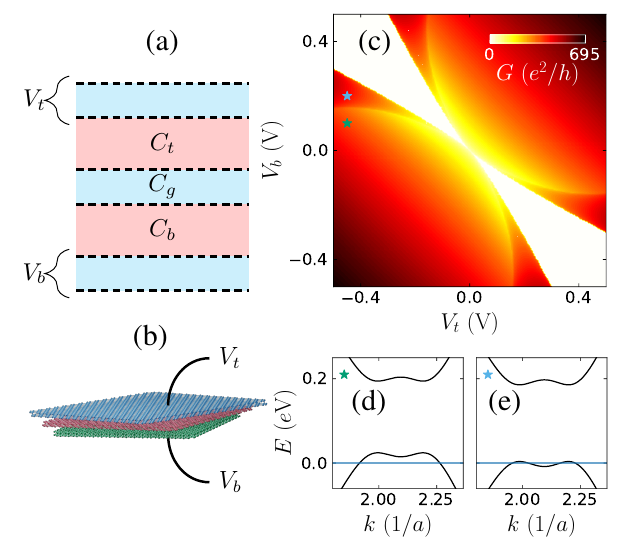}
 \caption{ \textbf{BLG dual gated by twisted top and bottom BLG.}
  (a) Side view of the device.
  (b) Cartoon illustrating the design.
  (c) Conductance of the middle BLG as a function of the top and bottom gate voltages obtained with the quantum capacitance model for an assumed sample width of 1 $\mu$m. 
  (d, e) Band structures at the voltage configurations marked by the stars in (b).
  } \label{fig:TBGgate}
\end{figure}

\paragraph*{Graphene-gated Bernal stacked bilayer graphene.}
Etched graphene can be used to define graphene side gates in planar graphene devices \cite{Molitor2007sideGates, Stampfer2008sideGates}. Here we propose a device based on stacked graphene layers, where the outer layers are used as gates.
The decoupling of large-angle twisted graphene layers can be employed to build a thin capacitor with an enormous geometrical capacitance. 
Thus, a small voltage applied to the top and bottom BLG is sufficient to create a large
displacement field between the BLG layers leading to a huge band gap 
that can be continuously tuned in a very large range, from zero to 200 meV for experimentally accessible parameters.

The proposed setup consists of three stacked BLG flakes, as shown schematically in figure~\ref{fig:TBGgate}(a), where the outer BLGs are twisted at a large angle relative to the middle one. This can be for example achieved by tearing BLG and stacking the pieces at a controlled angle \cite{Kim2016tearandstack, XLiu2020, He2021}. The topmost and bottom-most bilayers are then contacted separately [figure~\ref{fig:TBGgate}(b)]. 
We assume $C_\mathrm{t}=C_\mathrm{b}=3.5 \mathrm{\mu F cm^{-2}}$ \cite{Rickhaus2019tdblg}.
In principle it should also be possible to stack single layer graphene at an angle on top and bottom of the BLG, however such an approach is more challenging in practice. 

Figure \ref{fig:TBGgate}(c) shows the conductance of the middle BLG as a function of the top and bottom gate voltage. 
In the present setup the large potential difference leads to opening of a large band gap and a resulting insulating state. This produces a large region of zero conductance near the charge neutrality point.
In the map it is possible to spot a distinct conductance dip further from the charge neutrality point, which corresponds to the 'Mexican hat' structure at the band edge \cite{McCann2013}. 
A representative band structure in the regime with the Fermi energy within the Mexican-hat range is shown in figure~\ref{fig:TBGgate}(d), while shifting the Fermi energy out of this range corresponds to crossing the dip [figure~\ref{fig:TBGgate}(e)]. This crossing corresponds to a topological Lifshitz transition which was directly measured only at finite magnetic field in bilayer graphene \cite{Varlet2014Lifshitz}.

\paragraph*{Decoupled multi-layer graphene.}
The iterative process can also be applied to a system composed of more graphene layers, provided that each one is twisted by a large angle such that all the layers are electrically decoupled. Such systems have been realized experimentally \cite{Sprinkle2009, Mogera2015multilayer}. For more details on the carrier density in $n$-layer graphene see \cite{supplement}.

\section{Discussion}
In conclusion, we developed self-consistent methods for the electrostatic calculations for electronically decoupled graphene multilayers. We fabricated encapsulated twisted bilayer graphene samples, and performed low-temperature transport measurements and quantum transport simulations. For the twisted double bilayer graphene sample described in reference~\cite{Rickhaus2019tdblg} we model the electrostatics and transport. The theoretical and measured conductance show excellent agreement for both tBLG and tdBLG, confirming the accuracy of the model and the extraordinary decoupling between the atomically-close multilayers. Having confirmed the applicability of the quantum capacitance model for bilayer graphene, we also apply it to design a dual-gated BLG device that uses the decoupling mechanism to create a thin capacitor, which allows for the observation of a very large and entirely tuneable energy gap in BLG. 
Our results show that the decoupling in large-angle twisted graphene systems can exploited to investigate the band structure in BLG close to the band edge, which in practice includes also the trigonal warping effect that is theoretically captured when the skew interlayer hopping is included in the BLG Hamiltonian.
The decoupling can also be employed in decoupled multi-layer graphene, where the sheets retain the remarkable mobility of single layer graphene, while conducting in multiple parallel planes. Such high-mobility, ultra-thin devices open doors for various application in electronics.

The good agreement with the experimentally measured conductance for tBLG and tdBLG suggests their applicability is not limited to graphene-based devices, but also for a broad class of systems consisting of decoupled conducting layers. 
The self-consistent method is suitable to other materials hosting Dirac carriers, for example topological insulator surface states, as well as described by other band structures \cite{Ziegler2020hgte}, opening a new area of capacitively-coupled materials.

\ack
This work was supported by the Ministry of Science and Technology grant MOST 109-2811-M-006-544. 
KR acknowledges funding through  the Deutsche Forschungsgemeinschaft (DFG, German Research Foundation) Project-ID No. 314695032--SFB 1277 (subproject A07).
This research was supported in part by PL-Grid Infrastructure.

\section*{References}

\bibliography{decoupled_graphene}

\end{document}